\documentclass[superscriptaddress,onecolumn]{revtex4}
\usepackage{graphicx,epsfig}
\usepackage{amsmath}
\usepackage {amssymb}
\usepackage{multirow}
\usepackage{xfrac}
\usepackage{ulem}
\usepackage{float}
\usepackage{color}

\newcommand{\be}{\begin{eqnarray}}
\newcommand{\ee}{\end{eqnarray}}
\newcommand{\bea}{\begin{eqnarray}}

\newcommand{\eea}{\end{eqnarray}}

\def\beq{\begin{equation}}
\def\eeq{\end{equation}}
\def\beqn{\begin{eqnarray}}
\def\eeqn{\end{eqnarray}}
\def\ba{\begin{eqnarray}}
\def\ea{\end{eqnarray}}

\def\xprim2bar{\overline{x}^{\prime\prime}}

\def\beq{\begin{equation}}
\def\eeq{\end{equation}}

\def\p{\partial}
\def\pr{\prime}
\def\pp{{\prime\prime}}

\newcommand{\f}{\frac}

\def\A5{(A_5)_{\rm lat}}

\makeindex

\begin{document}

\title{Formation of Bound States in  Quintessence Alternative Theories}



\author{George Koutsoumbas}
	\email{kutsubas@central.ntua.gr}
	\affiliation{Physics Division, School of Applied Mathematical and Physical Sciences, National Technical University of Athens, 15780 Zografou Campus,
    Athens, Greece.}

\author{Andri Machattou}
	\email{andrimachattou@hotmail.com}
	\affiliation{Physics Division, School of Applied Mathematical and Physical Sciences, National Technical University of Athens, 15780 Zografou Campus,
    Athens, Greece.}
	
	\author{Eleftherios Papantonopoulos}
         \email{lpapa@central.ntua.gr}
	\affiliation{Physics Division, School of Applied Mathematical and Physical Sciences, National Technical University of Athens, 15780 Zografou Campus,
    Athens, Greece.}

\date{\today }

\begin{abstract}

We study RNAdS black holes in the presence of  quintessence matter. We find that the presence of quintessence influences significantly the metric function and the Hawking temperature. For this kind of black holes phase transitions are present. We spot the phase transition points using the corresponding heat capacities, as well as the Helmholtz free energy for the canonical thermodynamic ensemble. Calculating the Regge and Wheeler potential for scalar perturbations, we find out that no bound states are present for stable RNAdS black holes in the presence of  quintessence matter.

\end{abstract}


\maketitle
\flushbottom

\tableofcontents

\newpage

\section{Introduction}

Recent astronomical observations suggest that our Universe  is expanding at an accelerating  rate \cite{Riess1,  Perlmutter1999}. The presence of a cosmological constant cannot fully explain this expansion \cite{Carroll2001,Copeland2006}. The dark sector of our Universe may account for this expansion. Approximately $70\%$ of the energy density in the universe is in the form of an exotic, negative pressure component, dubbed dark energy. The observational bounds on the properties of dark energy have continued to tighten.  Taking $w$ to be the ratio of pressure to density for the dark energy $ w = p/\rho,$ recent observations dictate the range  $-1\le w \le -\f{1}{3}.$ The borderline case of $w=-1$ of quintessence mimics the cosmological constant term. Different dark energy cosmologies, reviewed, e.g. in \cite{Chen} provide a detailed discussion of the ${\rm \Lambda}$CDM model in particular and connect it to the  well-established observational tests, which constrain the current cosmic acceleration. 

The presence of a quintessence field to explain the Universe expansion modifies General Relativity and new modified Gravity theories were proposed in the presence of dynamical scalar fields \cite{Hellerman2001}.  In these theories black hole solutions were found in the presence of the quintessence matter field. One of the first black hole solutions in the presence of  quintessence was discussed in \cite{Kiselev:2002dx}. In this work new static spherically-symmetric exact solutions of Einstein equations with the quintessential matter surrounding a black hole have been discussed. Further studies in the presence of quintessence matter have been carried out, new  black hole solutions were obtained, and the impacts of the quintessence field on their thermal quantities, shadows, and quasi-normal modes have been discussed  \cite{Chen2008}. The presence of dark matter and quintessence fields outside the horizon of a black hole has an  impact on  their thermodynamic properties, such as the temperature, and introduces new critical phenomena. In \cite{Ghaderi2018} the quantum correction on thermodynamics of the Reissner-Nordstr\"om black hole in the presence of the quintessence matter associated with dark energy has been investigated.  These modifications could lead to generalizations of classical thermodynamic results making this area of study very interesting and leading to a connection between a black hole event horizon and its entropy \cite{Bekenstein1973, Hawking1974}.

The stability of black holes is a central issue in General Relativity and it has been studied for a long time starting with the pioneering works by Regge and Wheeler, \cite{Regge:1957a} and Zerilli \cite{nollert}. 
Their method consists of a decomposition of perturbations on a spherically symmetric background geometry into tensor harmonics. They found that it is possible to put the equations for the  oscillatory motion into the form of a single second-order Klein-Gordon-type differential equation with an effective potential. Then the study of these equations can give us important information on the stability of the black hole. The first metric that has been studied by Regge and Wheeler was the Schwarzschild metric. Another important tool to study the stability of black holes is to consider the quasi-normal modes (QNMs), in particular to calculate their quasi-normal frequencies (QNFs) (\cite{Regge:1957a}, \cite{nollert}). The calculation of QNMs and QNFs may give information about the stability of matter fields that perturb a region outside  of a black hole, which, however, are not considered to back-react on the metric. 

Bound states for  particles around black holes is a very interesting effect and a detailed study has been carried out in \cite{Grain:2007gn}. A similar study can be done by the investigation of  QNMs. In both cases  the goal is to find the characteristic complex frequencies for a scalar field, which propagates in a black hole background. The scalar field must satisfy the boundary conditions at the black hole horizon and at spatial infinity. However, the physical meanings of bound states and quasi-normal modes are very different. In \cite{Grain:2007gn} a systematic study was carried out of  quanta trapped between the finite AdS black hole potential barrier and the infinitely broad well, which works as a barrier preventing  the particles from reaching infinity. 
These bound states may have cosmological consequences related  to the primordial power spectrum  \cite{barrau3, Carr:2005zd}. Also the phenomenology of these bound states and their spectra should be  studied in detail and experiments associated with black holes should also take  these states into account. 

The formation and propagation of  bound states  in the vicinity of a Galileon black hole has been discussed in  \cite{Koutsoumbas:2018gbd}. The Galileon black holes are generated by a scalar field coupled to the curvature \cite{Kolyvaris:2011fk, Rinaldi:2012vy, Kolyvaris:2013zfa} and they are spherically symmetric  hairy black holes. Classically this coupling has been studied extensively; it strongly influences the propagation of the scalar field compared to a scalar field minimally coupled to gravity.  Also the stability of the background Galileon black hole has been studied and it has been found that bound states were trapped in a Regge and Wheeler potential well or penetrating the horizon of the Galileon black hole depending on the strength of the derivative coupling of a scalar field to curvature. The formation of quantum bound states requires that there is a radial potential containing a local well for given ranges of black hole parameters in the Klein-Gordon (KG)  equation.

Motivated by the work in \cite{Kiselev:2002dx} in which quintessential matter surrounding a background spherically-symmetric black hole interacts with it, dressing it with quintessential hair, we study the eventual formation of bound states formed outside the horizon of this hairy  black hole. Then considering a general form of the metric function we derive a KG-like equation, in which the Regge and Wheeler potential appears. We consider the thermodynamics of the solution and we discuss the eventual formation of bound states for various values of the quintessence parameter $w$ for the hairy black hole. We consider matter which does not back-react with the quintessence black hole and we study whether the Regge and Wheeler potential is affected by this matter outside the quintessence black hole.

Several aspects of this work have been discussed previously. Previous papers \cite{similar} cover various general aspects of our work: the thermodynamics of black holes is well understood and calculations of quasi-normal modes have been carried out. However, either no charged black holes have been covered, or models with no $\f{r^2}{L^2}$ term have been studied; in other works quasi-normal mode calculations have been done, but no mention has been made of bound states. Wherever there is an overlap with previous papers, we agree with their results. 

In summary, we study some aspects of the thermodynamics and examine the eventual formation of bound states outside the horizon of a background spherically-symmetric RNAdS black hole, which is dressed with  quintessential hair. We work with the black hole presented in \cite{Kiselev:2002dx} in which quintessential matter surrounding a background spherically-symmetric RNAdS black hole back-reacts dressing it with quintessential hair. We also find that, as the charge $Q$ of the black holes increases, the black hole temperature approaches zero on the horizon, in contrast with the case with $Q=0.$ We have also checked the thermodynamics of the model and spotted phase transitions when the horizon value $r_h$ and charge $Q$ varies. For small $Q$ there exist regions of the horizon value, where the black hole is thermodynamically unstable. For large $Q$ the black hole is stable for all horizons. Calculating the Regge and Wheeler potential we study the eventual bound states formation outside the horizon of the black hole; the result has been negative.

\section{Regge-Wheeler Potential}
\label{genregge}

In this section we will consider a fairly general metric and we will review the Regge-Wheeler potential, described in \cite{Regge:1957a} and  \cite{nollert}. We are studying only the perturbations of a test scalar field on a fixed background. We consider the metric:
\be ds^2 = -f(r) dt^2 + \f{dr^2}{f(r)} + r^2 d\Omega^2~.\ee This convention for the metric signature will be followed throughout this work. We set $G=c=\hbar=1.$ The generalized Klein-Gordon equation reads:
\begin{eqnarray}
 &-&\f{r^2}{f} \p_{tt} \Psi(t,r,\theta,\phi) + r^2 f \p_{rr} \Psi(t,r,\theta,\phi) + \left(r^2 f^\pr + 2 r f\right) \p_r \Psi(t,r,\theta,\phi) \notag\\
&&+ \left[\f{1}{\sin\theta} \p_\theta(\sin\theta \p_\theta) + \f{1}{\sin^2\theta} \p_{\phi\phi} \right]\Psi(t,r,\theta,\phi) = 0~. 
\end{eqnarray}
We decompose $\Psi(t,r,\theta,\phi)$ into radial, angular and time parts, using the  ansatz
\be \Psi(t,r,\theta,\phi) = R(r) Y_{lm}(\theta,\phi)  e^{-i \omega t}~,\ee 
where $Y_{lm}(\theta,\phi)$ are the usual spherical harmonics, which satisfy the equation
\be \left[\f{1}{\sin\theta} \p_\theta(\sin\theta \p_\theta) + \f{1}{\sin^2\theta} \p_{\phi\phi} + l (l+1) \right] Y_{lm}(\theta,\phi)=0~.\ee
The radial part of the wave function will then satisfy:
\be (r^2 f R^\pr )^\pr +\left(\f{\omega^2 r^2}{f} - l (l+1) \right) R = 0.\ee 
We define \be R\equiv \f{u}{r}~,\label{udef}\ee
and the equation becomes:
\be f^2 \f{d^2 u}{d r^2} +  f \f{d f}{d r} \f{d u}{d r} + \omega^2 u -\f{l (l+1) f}{r^2} u - \f{f}{r} \f{d f}{d r} u = 0. \ee
Using the tortoise coordinate
\be \f{d}{d r} =\f{1}{f} \f{d}{d r_*}~,\ee
we get
\be \f{d u}{d r} =\f{1}{f} \f{d u}{d r_*},\  \f{d f}{d r} =\f{1}{f} \f{d f}{d r_*},\ \f{d^2 u}{d r^2} = \f{1}{f^2} \f{d^2 u}{d r_*^2} - \f{1}{f^3} \f{d f}{d r_*} \f{d u}{d r_*}~,\ee 
and the radial equation takes a Klein-Gordon-like form.
 \be -\f{d^2 u}{d r_*^2} + V u = \omega^2 u~,\label{radeq}\ee 
where the Regge-Wheeler potential reads:
\be V =f\left[ \f{l(l+1)}{r^2}+\f{1}{r}\f{d f}{d r}\right].\ee

\section{Quintessence Matter and its Thermodynamics} 
\label{quint}

In this section we introduce quintessence matter outside the horizon of a black hole and we study the thermodynamics of the coupled system.

\subsection{Description of Quintessence} 

As we have already discussed, quintessence is a scalar field, whose equation of state parameter $w$ is defined as the ratio of its pressure $p$ and its energy
density $\rho$ and is given by \be w=\f{p}{\rho},\ee where $-1\le w \le -\f{1}{3}$ for homogeneous quintessence field.
Writing the line element in the form: 
\be ds^2 = -f dt^2 + \f{dr^2}{f} + r^2 d\Omega^2,\ee 
it turns out that there exist two solutions for the Einstein equations, the well-known Schwarzschild solution and a novel one \be f=1+\f{a}{r^{1+3 w}},\ee where $a$ and $w$ are constants. The Einstein equations yield:
\be \rho = \f{a}{2}\f{3 w}{r^{3 (w+1)}},\ R = 3 a w \f{1-3 w}{r^{3(1+w)}}~.\ee 
Additivity and linearity yield \cite{Kiselev:2002dx} the solution for the metric: \be f = 1 -\f{2 M}{r} + \f{Q^2}{r^2} + \f{r^2}{L^2} - \f{a}{r^{1+3 w}}~.\ee The metric function in the AdS black hole, apart from the usual dependence on the black hole mass $M$ and charge $Q,$ also depends on the parameters $a$ and $w,$ which play a crucial role in the thermodynamics of the system and the formation of bound states. In addition the AdS radius $L$ enters the metric.

\subsection{Thermodynamics in AdS Spacetime}

The mass can be expressed in terms of the remaining parameters and the horizon radius $r_h,$ since $f(r_h)=0:$ 
\be M = \f{r_h}{2}\left[1+\f{Q^2}{r_h^2} + \f{r_h^2}{L^2} - \f{a}{r_h^{1+3 w}}\right]~.\ee 
The Hawking temperature is given by:
\be T = \f{1}{4 \pi} f^\pr(r_h) = \f{1}{4 \pi} \left[\f{2 M}{r_h^2} - \f{2 Q^2}{r_h^3} + \f{2 r_h}{L^2} + \f{a (1+3 w)}{r_h^{2+3 w}}\right]~, \ee
while the entropy is given by \be S=\int\f{dM}{T} = \f{A}{4} = \pi r_h^2~. \ee 
In the following we will investigate how the above quantities are affected by the quintessence parameters $a$ and $w.$ From now on we measure lengths in units of $L,$ that is, we set $L=1.$

\begin{figure}[ht]
\begin{center}
\includegraphics[scale=0.8,angle=0]{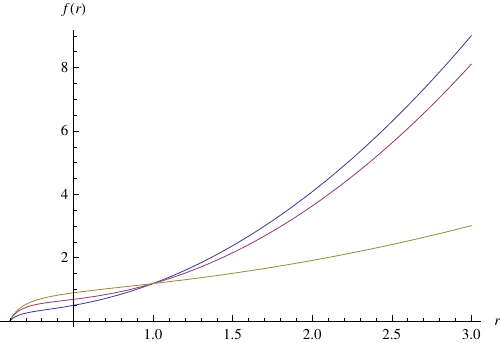}
\includegraphics[scale=0.8,angle=0]{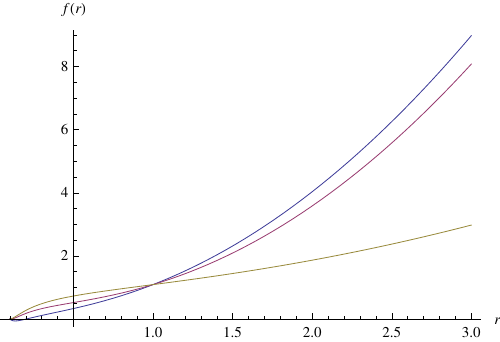}
\end{center}
\caption {(a) Metric function $f$ versus $r$ for $Q=0.00,$ $a=0.80,\ r_h=0.10$ and $w=-0.40$ (blue curve, highest) $w=-0.60$ (purple curve, middle) and $w=-1.00$ (khaki curve, lowest). (b) Metric function $f(r)$ versus $r$ at $Q=0.10,$ $a=0.80,\ r_h=0.10$ and $w=-0.40$ (blue curve, highest) $w=-0.60$ (purple curve, middle) and $w=-1.00$ (khaki curve, lowest).} \label{metric}
\end{figure}

In Fig. \ref{metric}, left panel, we plot the metric function $f(r)$ for $Q=0.00,\ a=0.80$ and three values of the $w$ parameter. The $w$ is modeled after the cosmological parameter, for which \cite{Chiba:1999wt} gives the bounds $-1<w<-\f{1}{3},$ but this refers to homogeneous dark energy; in this paper we deal with an anisotropic fluid around a black hole, so that the bounds may be different. To explore the model, however, we have chosen values of $w$ lying within the above bounds; in particular we have selected for $w$ the values $-0.40,\ -0.60$ and $w=-1.00.$ The metric function becomes small on the right hand side of the diagram, as $|w|$ increases, and takes its lowest values for $w=-1.00$ In the right panel similar plots for $Q=0.10$ are depicted: there are minor quantitative differences from the case in the left panel. 

\begin{figure}[ht]
\begin{center}
\includegraphics[scale=0.8,angle=0]{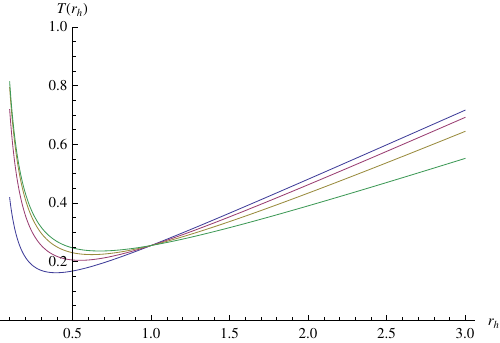}
\includegraphics[scale=0.8,angle=0]{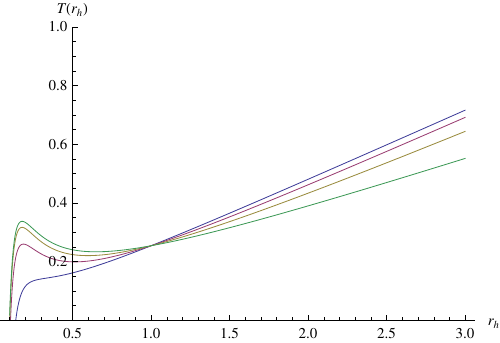}
\end{center}
\caption {(a) Temperature $T$ versus $r$ for $Q=0.00,$ $a=0.80,\ r_h=0.1$ and $w=-0.40$ (blue curve, highest on the right part) $w=-0.60$ (purple curve, middle), $w=-0.80$ (khaki curve), $w=-1.00$ (cyan curve, lowest). (b) Temperature $T$ versus $r$ at $Q=0.10,$ $a=0.80$ $r_h=0.1$ and $w=-0.40$ (blue curve, monotonic) $w=-0.60$ (purple curve, middle), $w=-0.80$ (khaki curve), $w=-1.00$ (cyan curve, lowest).} \label{temperature}
\end{figure}

In Fig. \ref{temperature}, left panel, we plot the temperature versus $r_h$ for $Q=0.00,\ a=0.80$ and various values of the $w$ parameter. The temperature decreases for increasing $|w|.$ Moreover the temperature is not a monotonic function and the "small" and "large" black holes (depending on the size of the horizon $r_h$ in units of the AdS radius $L$) respond differently to the quintessence. The temperature has a minimum at some value of the horizon and this may signal a phase transition as one moves from "small" to "large" black holes. We will elaborate on this behaviour in the sequel. 

For $Q>0,$ a major difference from the $Q=0$ case is that now the temperature vanishes at the horizon. For $Q=0.10$ and $w=-0.40$ the temperature rises monotonically and its behaviour is not very interesting. On the contrary, for $w=-0.60$ and $w=-1.00$ the temperature has a local maximum and then a local minimum, so the black hole is expected to have two phase transitions as $r_h$ varies. We will see that this is exactly the case. It turns out that no phase transition exists beyond some value of $Q.$ We will examine these issues in the following subsection.

\subsection{Heat Capacity}

\begin{figure}[ht]
\begin{center}
\includegraphics[scale=0.8,angle=0]{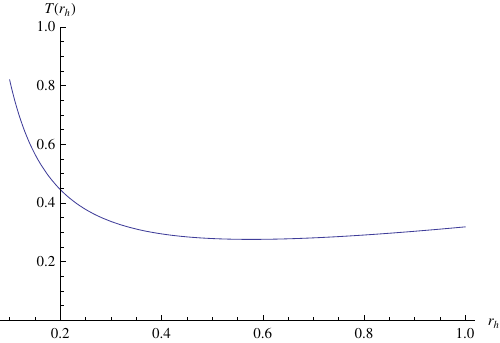}
\includegraphics[scale=0.8,angle=0]{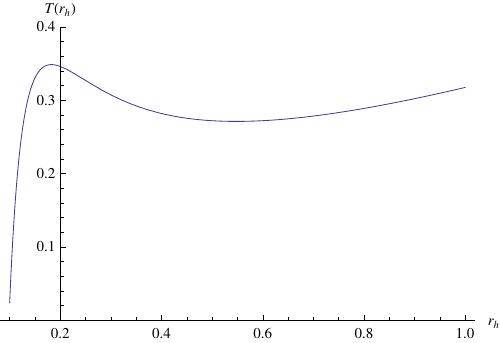}
\includegraphics[scale=0.8,angle=0]{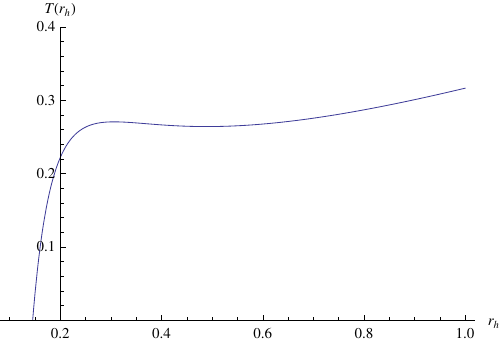}
\includegraphics[scale=0.8,angle=0]{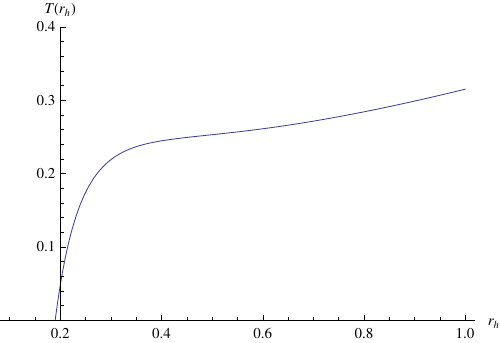}
\end{center}
\caption {Temperature $T$ for $a=0$ and $Q=0.00,\ 0.10,\ 0.15,\ 0.20.$} \label{T}
\end{figure}

We would like to elaborate on the meaning of the extrema of the temperature as a function of $r_h.$ Parameters such as the charge $Q,$ the cosmological constant, and the quintessence quantities $a$ and $w$ are held constant, while $r_h$ varies. The behaviour of the heat capacity will be used to shed light on possible phase transitions. We start with the model in which we set $a=0.$ Figure \ref{T} depicts the temperature of the black hole at four values of the charge: $Q=0.00, \ Q=0.10,$ (upper two panels) and $Q=0.15$ and $Q=0.20$ (lower two panels). It is evident that, for $Q=0.00,$ the slope of the temperature curve changes sign around $r_h=0.57,$ while, for $Q=0.10,$ it changes slope sign twice, at $r_{h1}=0.18$ and $r_{h2}=0.55.$ These two values for $r_h$ move to the values $r_{h1}=0.32$ and $r_{h2}=0.48$ for $Q=0.15,$ while, at $Q=0.20,$ there is no sign flip any more; the temperature increases monotonically with $r_h.$  

\begin{figure}[ht]
\begin{center}
\includegraphics[scale=0.8,angle=0]{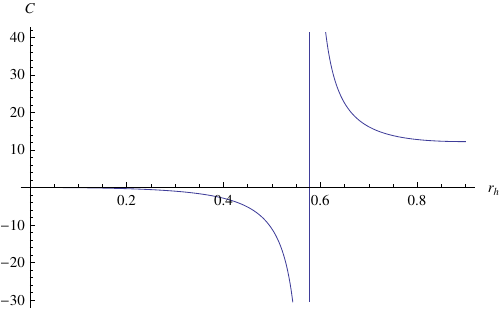}
\includegraphics[scale=0.8,angle=0]{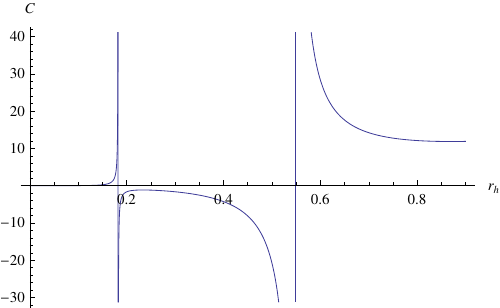}
\includegraphics[scale=0.8,angle=0]{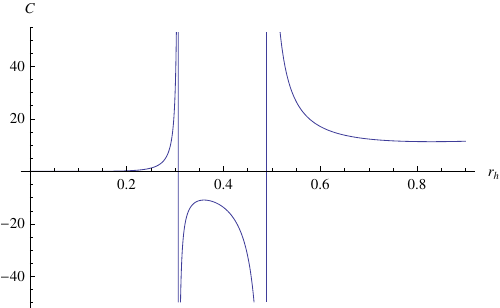}
\includegraphics[scale=0.8,angle=0]{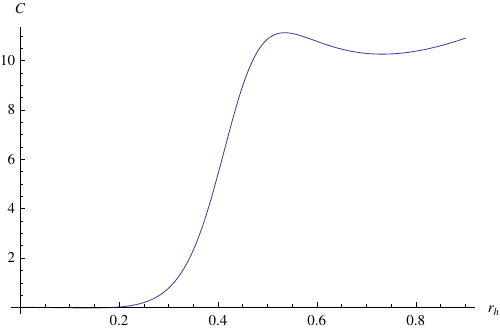}
\end{center}
\caption {C for $a=0$ and $Q=0.00,\ 0.10,\ 0.15,\ 0.20$ versus $r_h.$} \label{C}
\end{figure}

To understand this behaviour better, we calculated the specific heat \be C=\f{d M}{d T} = \f{\f{d M}{d r_h}}{\f{d T}{d r_h}}\ee at the same values of the parameters as before. The results are depicted in figure \ref{C}. For small values of $r_h$ we find the well-known negative sign for the specific heat, when $Q=0,$ but the heat capacity turns positive at $r_h=0.57.$ This sign change may be associated  with a phase transition from small towards large black holes. Things change, as soon as $Q$ takes non-zero values and implies additional phase transitions at the two values of $r_{h1}$ and $r_{h2}$ referred to above. As one moves to larger values of $Q,$ the two phase transitions approach each other, until, for $Q=0.20,$ there is no phase transition at all and the system is stable for all values of $r_h.$ One may also describe this behaviour alternatively: starting with large $Q$ values, no phase transition is expected to exist and the system is stable. If we move towards smaller values of $Q,$ phase transitions appear at two values of $r_h,$ the ones we called $r_{h1}$ and $r_{h2}.$ Between these two values the system is unstable, while, otherwise, it is stable. For even smaller values of $Q$ the phase transition at $r_{h2}$ moves towards $r_h=0.57,$ while the phase transition at $r_{h1}$ moves towards $r_h=0$ and it finally disappears. The system at $Q=0$ is unstable for small values of $r_h$ and stable for the remaining values. We have checked that there is no slope flip in the $M$ versus $r_h$ curves, so this behaviour is due to the variation of the temperature, in particular to the sign change in its slope: We will see how this phase structure changes under the influence of quintessence. 

Let us summarize the results for the $a=0$ cases. Figure \ref{C} strongly suggests that, in some intervals of $Q,$ there are two values of $r_h,$ denoted by $r_{h1}$ and $r_{h2} (>r_{h1}),$ for which there is a sign change for the heat capacity, signaling a phase transition. For $0<r_h<r_{h1}$ the heat capacity is positive, for $r_{h1}<r_h<r_{h2}$ it is negative and for $r_h>r_{h2}$ it is positive once more. In regions where the heat capacity is negative the black hole is thermodynamically unstable; this means that, if it is put in contact with a heat bath with the same temperature, it will either absorb energy and cool further, or lose energy and heat up further. None of these scenarios permit stable equilibrium. On the contrary, when the heat capacity is positive, a stable thermal equilibrium is possible. Positive heat capacity has previously been found in Kerr - Newman black holes, as well as in AdS black holes. 

\begin{figure}[ht]
\begin{center}
\includegraphics[scale=1.0,angle=0]{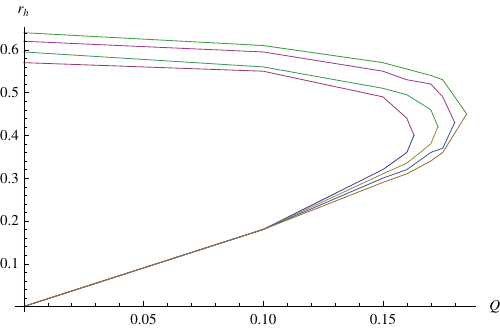}
\includegraphics[scale=1.0,angle=0]{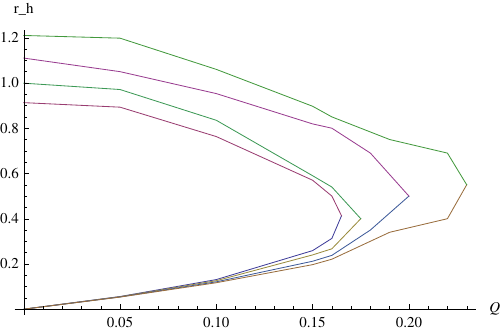}
\end{center}
\caption {(a) Values of $r_h,$ where sign change occurs for $C,$ at $w=-0.80$ and $a=0.0,\ 0.30,\ 0.70,\ 1.00.$ Notice that $a=0.0$ is the innermost curve. (b) Points of flip from small to large black holes, according to the entropic diagram. Parameters are the same, as in the left  panel.} \label{various_a}
\end{figure}

In figure \ref{various_a}, left panel, we depict the values $r_{h1}$ and $r_{h2}$ for $a=0,\ 0.30,\ 0.70,\ 1.00;$ $w$ has been set equal to $-0.80.$ The $a=0$ case corresponds to the innermost curve. We observe that $r_{h1}$ vanishes when $Q=0,$ and the black hole has only one phase transition point, at the value $r_h=r_{h2}.$ In this case the black hole starts with negative heat capacity, which changes to positive values at $r_h=r_{h2}.$ For $Q>0$ both values $r_{h1}$ and $r_{h2}$ are present. The two values $r_{h1}$ and $r_{h2}$ approach each other when $Q$ increases and, at some critical value for $Q,$ they coincide and then there is no more phase transition. The heat capacity is everywhere positive and the charged black holes are stable when their charge is large enough. In view of this behaviour, the search for possible bound states must be limited in the regions, where the black holes are stable. The region in the interior of the curve for each $a$ in figure \ref{various_a}, is the region of instability.

\begin{figure}[ht]
\begin{center}
\includegraphics[scale=1.0,angle=0]{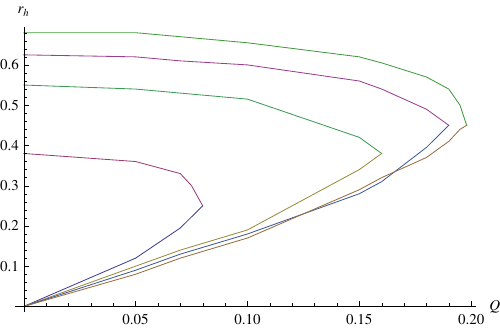}
\includegraphics[scale=1.0,angle=0]{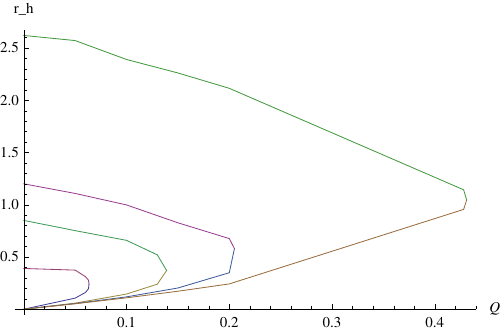}
\end{center}
\caption {(a) Values of $r_h,$ where sign change occurs for $C,$ at $a=0.85$ and $w=-0.40,\ -0.60,\ -0.80,\ -1.00.$ Notice that $w=-0.40$ is the innermost curve. (b) Points of flip from small to large black holes, according to the entropic diagram. Parameters are the same, as in the left  panel.} \label{various_w}
\end{figure}

In figure \ref{various_w}, left panel, one may find a similar graph, depicting regions of stability and instability in the $r_h-Q$ plane for $a$  fixed to the value $0.85$ and various values of $w,$ namely $w= -0.40,\ -0.60,\ -0.80,\ -1.00.$ The results show that, the smallest $|w|$ is, the smallest the instability region appears to be.

\begin{figure}[ht]
\begin{center}
\includegraphics[scale=1.9,angle=0]{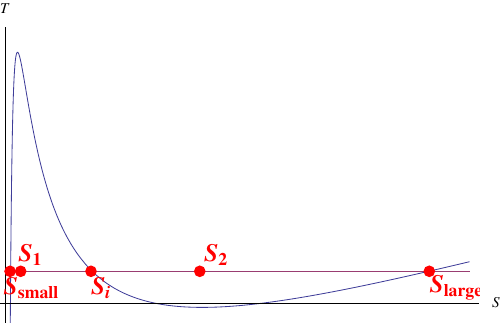}
\end{center}
\caption {Points in the $T-S.$ The straight line above the horizontal axis is the curve $T=T_*.$} \label{points_therm}
\end{figure}

The previous results on the sign change of the heat capacity correspond to the question of the local thermodynamic stability. However, this is not the whole story, since the true phase transition is related with the global thermodynamic considerations. In the latter approach, one calculates the Helmholtz free energy \be F(r_h) = M(r_h) - T(r_h) S(r_h).\ee This will permit us to study the small-versus-large black holes transition. We will use the canonical ensemble, in which the charge $Q$ is held fixed. This choice is connected with the small to large black hole hole transition. The Helmholtz free energy function $F(r_h)$ is smooth; however, for some values of the charge $Q,$ the function $T(r_h)$ is not monotonic; several values of $r_h$ correspond to the same temperature, yielding different small and large black-hole entropies. As may be seen in figure \ref{points_therm}, the horizontal line at $T=T_*,$ for some temperature $T_*$ to be determined, may intersect the curve $T(S)$ in three points, namely $S_{small} < S_{i} < S_{large},$ a small, an intermediate and a large value. These values of the entropy correspond to three values for the r-coordinate, via $S_{small}=\pi r_{small}^2,\  S_{i}=\pi r_{i}^2,\  S_{large}=\pi r_{large}^2.$ Points $S_1$ and $S_2$ in the figure represent the extrema of the curve. These are the points determined by the extrema of the temperature and related to the sign change in the heat capacity that we found previously. The small to large transition occurs when $F_{large}(T_*) = F_{small}(T_*).$ At this temperature the black hole may jump (discontinuously) from small to large (entropy, as well as) horizon radius. The values of the entropy and the horizon radius may be found, once the value $T_*$ is determined, via a Maxwell equal-area construction in the $T–S$ plane. 

The Helmholtz free energy function $F(r_h) = M(r_h) - T(r_h) S(r_h)$ for the canonical ensemble (with fixed charge) yields: \be d F = dM - T dS - S dT = - S dT \Rightarrow F(T) = F(T_0) -\int_{T_0}^T S d T.\ee Integrating by parts: 
\be F(T) = F(T_0) - S T +S_0 T_0 + \int_{S_0}^S T d S \ee $T_0$ is a reference value and $F(T_0)$ the corresponding value of the free energy.
A phase transition occurs when the free energies of the small and large black holes are equal: 
\be F_{large}(T_*)=F_{small}(T_*)\ee 
\be \Rightarrow F(T_0) - S_{large} T_{large} + S_0 T_0 + \int_{S_0}^{S_{large}} T d S = F(T_0) - S_{small} T_{small} + S_0 T_0 + \int_{S_0}^{S_{small}} T d S\ee 
\be \Rightarrow   S_{large} T_{large} - S_{small} T_{small} = \int_{S_{small}}^{S_{large}}  T d S  \Rightarrow T_*[S_{large}-S_{small}] = \int_{S_{small}}^{S_{large}}  T d S,\ee 
since $T(r_{large})=T(r_{small})=T_*.$ 

Rearrangement of the last relation yields: \be \int_{S_{small}}^{S_{large}} (T(S) - T_*) d S=0,\ee which is the equal-area condition. The temperature $T_*$ is tuned, so that the above equality is satisfied. Then it is straightforward to locate $r_{small},\ S_{small}$ and $r_{large},\ S_{large}.$ 

The extrema of the function $T(S)$ are determined by the equations: $\f{d T}{d S} = 0 \Rightarrow \f{d T}{d r_h}=0,$ 
since $S=\pi r_h^2,$ so they correspond to the values of $r_h,$ where the heat capacity changes sign and it follows that $r_{small} < r_{h1} < r_{h2} < r_{large}.$ Thus the interval of instability is slightly wider than the one estimated via the heat capacity data.

In figure \ref{various_a}, right panel, one may find the results at $w=-0.80$ and $a=0.0,\ 0.30,\ 0.70,\ 1.00,$ obtained through the use of the Helmholtz free energy. The parameters used are the same as the ones used in the left panel of this same figure, to permit a comparison of the two approaches. As explained above, the inequalities $r_{small} < r_{h1} < r_{h2} < r_{large}$ hold, suggesting that the contours in the right panel should define wider instability regions than the corresponding contours in the left panel. This is indeed the case, as depicted in the right panel.

A similar approach as the one that lead to figure \ref{various_a}, right panel, yields the diagram in figure \ref{various_w}, right panel. Parameter $a$ is $0.85$ in this figure, while the values of $w$ are $-0.40$ (the innermost curve), $-0.60,\ -0.80$ and $-1.00$ (the outer curve). It is obvious that, for each value of $w,$ there exist one or  two values of $r_h,$ the ones that we called $r_{h1}$ and $r_{h2}$ previously. Between these values the black holes are unstable. These values depend on $Q$ as well and we draw a figure similar to \ref{various_a}, right panel, to state the results clearly. In figure \ref{various_w} the points between which the heat capacity is negative are shown. The black holes are unstable in the interior part of these curves. It is evident that there exist limiting values for $Q,$ beyond which there is no instability. Once more, the instability regions in the right panel of figure \ref{various_w} are wider than the ones depicted in the left panel of the same figure.

\section{Calculation of Bound States}
\label{BS}

We recall the metric in the AdS spacetime with quintessence
\be ds^2 =-f(r) dt^2+\f{dr^2}{f(r)}+r^2 d\Omega^2,\ \ \ f(r)=1-\f{2 M}{r} + \f{Q^2}{r^2} + r^2 - \f{a}{r^{1+3 w}}~.\label{line}\ee 
We concentrate specifically on the potential for scalar perturbations.
The Regge-Wheeler potential $V$ for scalar perturbations reads:
 \be V = f(r)\left(\f{f^\pr(r)}{r}+\f{l (l+1)}{r^2}\right)~.\label{Vscalar}\ee  One should also state the boundary conditions, which define the bound states. Near the horizon the function $f(r)$ tends to zero and the physical boundary condition reads: \be u(r) \approx e^{-i \omega r_*},\ee where the function $u(r)$ has been defined via equation (\ref{udef}), where $r_*$ is the tortoise coordinate. This means that the excitations are purely in-going at the horizon. At infinity, the radial equation (\ref{radeq}) for $u(r)$ simplifies to: \be r^2 u^\pp + 4 r u^\pr \simeq 0.\ee The solution of this equation is given by the expression: \be u(r) \simeq A + \f{B}{r^3},\ee where $A$ and $B$ are the integration constants. We choose the Dirichlet boundary conditions, which are defined through $\lim_{r\to+\infty} u(r)=0\Rightarrow A=0.$ 

\begin{figure}[ht]
\begin{center}
\includegraphics[scale=1.3,angle=0]{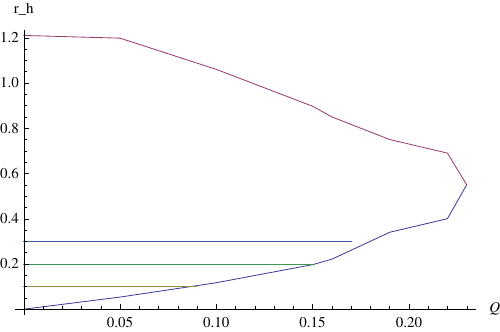}
\end{center}
\caption {Ranges of $Q,$ which may give bound states, if $w=-0.80$ and $a=1.0.$} \label{bound}
\end{figure}

We have tried to spot bound states using the method described above, using the values $a=1.0,\ w=-0.80.$ We reproduce the relevant part of figure \ref{various_a} as figure \ref{bound}. It has turned out that the potentials obtained may have problems, which do not permit the existence of bound states; for example, they may be too shallow, or have more than one minima. We have indicated in figure \ref{bound} the ranges, for which the potentials might in principle support bound states. The estimates are semi-quantitative. The potentials in these range of charges $Q$ are not guaranteed to support bound states; they are just not excluded from the beginning. The intervals, where bound states are possible are even smaller than the ones indicated in figure \ref{bound}. The $Q$ ranges for potentials, which are not rejected in principle, are indicated in figure \ref{bound} for $r_h=0.10,\ r_h=0.20$ and $r_h=0.30.$ We observe that the allowed intervals of $Q$ in the sense described above lie totally in the thermodynamically unstable region. For larger values of $r_h$ we have found no possibility of bound states at all, since the potentials turn out to be too flat. Thus we conclude that there exist no bound states in the thermodynamically stable region of these models. These and other similar problems, are not alleviated by changing parameters, such as using larger angular momentum numbers $l.$ 

\section{Conclusions}
\label{conclusions}

We have studied black hole solutions in the presence of quintessence matter. It has been shown that the parameters of the quintessence influence significantly the metric function and the Hawking temperature. It has turned out (Fig. \ref{temperature}) that, for $w=-0.40,$ "large" black holes $(r\gtrsim 0.6 L)$ have larger temperatures and radiate more than "small" ones, while, as $|w|$ increases, the behaviour changes: for $w=-1.00,$ "small" black holes have larger temperatures and radiate more than "large" ones.  "Large" black holes with large $|w|$ are expected to be longer lived, compared to "large" black holes with small $|w|,$  The situation is reversed for small black holes. In the long run, one may expect to find mainly large black holes with large $|w|,$ or small black holes with small $|w|,$ if other factors (e.g. angular momentum or charge) do not change. 

We have examined the thermodynamics of the black hole with quintessence fields. We began with SAdS black hole (with $a=0$) and found that, for large values of $Q,$ the heat capacity was positive, suggesting a thermodynamically stable black hole. When $Q$ took on smaller values, an interval $(r_{h1},\ r_{h2})$ in $r_h$ appeared, where the heat capacity was negative, while in the remaining regions of $r_h$ the heat capacity remained positive. Phase transitions are expected between the various phases. For still smaller values of $Q$ (including the $Q=0$ case) the interval $(r_{h1},\ r_{h2})$ became broader. At $Q=0$ the bound $r_{h1}$ was equal to zero, so that only one phase transition remained. In the cases, where $a > 0,$ the picture was qualitatively similar. Since the heat capacity detects the local thermodynamic stability, we have also calculated the Helmholtz free energy and determined the instability regions once more. The results are depicted in the right hand panels of figures \ref{various_a} and \ref{various_w}; the instability regions found in this way are larger than the ones found considering the heat capacities.

Finally we searched for possible formation of bound states varying the various parameters. We have found that no bound states can be found for thermodynamically stable black holes; it is possible to find bound states within the instability regions, so that the thermodynamic stability and the bound state formation appear to be incompatible.

\vspace{1.0cm}

{\bf Acknowledgement}: We thank the anonymous referee for constructive criticism and very helpful suggestions.

\end{document}